\documentclass[10pt,twocolumn,twoside]{IEEEtran}

\usepackage{graphicx}
\usepackage{verbatim}
\usepackage[cmex10]{amsmath}
\usepackage{amssymb}
\usepackage{amsfonts}
\usepackage{bbm}
\usepackage{hyperref}
\usepackage{color}
\usepackage{float}

\usepackage{algorithmic}
\usepackage{array}
\usepackage{multirow}
\ifCLASSOPTIONcompsoc
 \usepackage[caption=false,font=normalsize,labelfont=sf,textfont=sf]{subfig}
\else
 \usepackage[caption=false,font=footnotesize]{subfig}
\fi
\usepackage{url}
\usepackage[noadjust]{cite}

\begin{document}

\title{Replay Spoofing Countermeasure Using Autoencoder and Siamese Network on ASVspoof 2019 Challenge}

\author{{Mohammad Adiban, Hossein Sameti, Saeedreza Shehnepoor}
\thanks{
%Lu Chen, Zhi Chen, Bowen Tan, Sishan Long, and Kai Yu are supported by the National Key Research and Development Program of China under Grant No.2017YFB1002102, and Shanghai International Science and Technology Cooperation Fund (No. 16550720300). Milica Ga{\v{s}}i{\'c} is supported by an Alexander von Humboldt Sofja Kovalevskaja award. 
% (Corresponding author: Kai Yu)
The authors are with the Department of  Computer Engineering, Sharif University of Technology, Tehran, Iran (e-mail: adiban@ce.sharif.edu; sameti@sharif.edu; shehnepoor@ce.sharif.edu).

}}

%\markboth{IEEE/ACM TRANSACTIONS ON AUDIO, SPEECH, AND LANGUAGE PROCESSING,~Vol.~0, No.~0, December~0}
%{Chen \MakeLowercase{\textit{et al.}}: AgentGraph: Towards Universal Dialogue Management with Structured Deep Reinforcement Learning}

\maketitle

% As a general rule, do not put math, special symbols or s
\begin{abstract}

Automatic Speaker Verification (ASV) is the process of identifying a person based on the voice presented to a system. Different synthetic approaches allow spoofing to deceive ASV systems (ASVs), whether using techniques to imitate a voice or recunstruct the features. Attackers try to beat up the ASVs using four general techniques; impersonation, speech synthesis, voice conversion, and replay. The last technique is considered as a common and high potential tool for spoofing purposes since replay attacks are more accessible and require no technical knowledge from adversaries. In this study, we introduce a novel replay spoofing countermeasure for ASVs. Accordingly, we used the Constant Q Cepstral Coefficient (CQCC) features fed into an autoencoder to attain more informative features and to consider the noise information of spoofed utterances for discrimination purpose. Finally, different configurations of the Siamese network were used for the first time in this context for classification. The experiments performed on ASVspoof challenge 2019 dataset using Equal Error Rate (EER) and Tandem Detection Cost Function (t-DCF) as evaluation metrics show that the proposed system improved the results over the baseline by 10.73\% and 0.2344 in terms of EER and t-DCF, respectively.
\end{abstract}


\IEEEpeerreviewmaketitle

\section{Introduction}
\label{sec:introduction}
Speech can be considered as one of the most important means of communication for the human. Each individual has a unique voice pattern, identifiable as a signature. This pattern helps people to identify the other communication end. Consequently, tremendous low-cost technologies have been developed based on voice as a biometric feature for identity recognition \cite{Jain2006} known as “Automatic Speaker Verification” systems (ASVs). ASV captures different clues such as intonations and vocal tract shapes in order to verify a person’s identity \cite{Kinnunen2010}. On the other hand, there are techniques available to synthesize the voice or its characteristics \cite{Amin2014}\cite{Masuko2000}. This provides a great opportunity for spoofing, where the attacker exploits a specific speaker's voice to spoof an ASV system. Besides, advances in the channel and noise detection and removal of their effects have made the ASV application suitable for market usage. The issue become severe when it comes to e-commerce\footnote{\href{http://www.nuance.com/ landing-pages/products/voicebiometrics/vocalpassword.asp}{http://www.nuance.com/ landing-pages/products/voicebiometrics/\\vocalpassword.asp}} and smartphone logical access scenarios \cite{Lee2013}. These ASVs mostly use short-term spectral features which are very vulnerable to spoofing attacks. Accordingly, the research effort mainly investigates four types of attacks to address this problem: impersonation, speech synthesis, voice conversion, and replay. By mimicking the human voice or even alter that, attackers deceive the ASVs, for the first category . Formant extraction ($F_0, F_1, and F_2$) is a regular feature based approach to spot spoof in this category \cite{Amin2014}\cite{Lau2004}\cite{Eriksson1997}. The second category focuses on Text to Speech (TTS) approaches where a system synthesizes a text to generate the voice. In the third category, the voice of a person is converted from one person to the target speaker and then it is presented to ASVs for spoofing \cite{Stylianou2009}\cite{Evans2015}. Finally, attackers take use of recorded voices from the genuine speakers for spoofing, to form a replay attack.
Several studies extract features from speech and classify spoof/genuine utterances \cite{Lindberg1999}\cite{Villalba2010}.
However, there are problems still remaining out of attention. As \cite{Kaavya2016} mentioned, one of the important aspects to consider is determining which features or classifiers are better for discrimination. Furthermore, transforming features into space with a more informative context for discrimination is also an interesting research direction should be considered. Here in this work, we employ the Siamese networks for the first time as a classifier to increase the level of discrimination strength between the extracted features. Although Siamese networks were used for spoofing detection tasks such as a discrimiative feature extraction \cite{Kaavya2018}, the performance of such networks as a classifier is overlooked in ASVs. Our motivation for using such a method is owing to the fact that Siamese networks are applicable to tasks including measuring similarity or determining relationships between two comparable subjects \cite{bell2015learning}. In such tasks, two identical subnetworks are usually used to process the two injected inputs, and another module takes the outputs of these two sub-networks and makes the final decision. Siamese networks can be fed by two sets of features and after taking the final output from the output layer, they will use a distance measure in order to spot an input as a spoof/non-spoof sample.\\
In this study, we also extract CQCCs as basic features from both genuine and spoof samples. Since these features are sparse and raw, we feed them to autoencoders to have compressed and more discriminative features. In particular, our motivation for using autoencoder in this context is due to the fact that replay speech suffers from channel or convolutional noise, since the recording process is performed by two microphones, through one loudspeaker \cite{Adiban2017}. As a result, noise detection is one of the other aspects of replay spoofing detection as it is one of the most important factors in the successful adaptation of ASVs in the market place. For instance, in \cite{Yin2015}, a robust system is developed to improve Deep Neural Network (DNN) performance by using noisy data from channels. This type of systems is called “Noise Aware” systems, where noise is introduced to the system, in order to help the system to detect spoofing attacks especially for replay ones. In this study, Autoencoder Neural Netowork is employed, due to their prior success in extracting the noise as latent variable, given the same noisy input as the network's output \cite{Sun2016}\cite{hsu2017unsupervised}. We trained the autoencoder in order to  consider the noise and extract an informative representation of basic features, capable of representing both meaningful and sparse features related to original samples. Moreover, in contrast with other feature reduction and selection methods no feature is excluded, Hence, for both clean and noisy speech all features are preserved. Finally, in order to have contextual representation of features, we first feed them to Convolutional Neural Network (CNN) and then share the weights of CNN with another CNN identical to the first CNN and afterwards, they are employed in Siamese network. Siamese networks extract the similarity or a connection between two inputs and result in a significant discriminative classification.\\
In the following, first we list the related studies in Section \ref{sec:related-works}, then we will define the models of our works in Section \ref{sec:models}. Subsequently, the proposed system is introduced in Section \ref{sec:proposed-system}. Next, we give a brief explanation of the experimental setup in Section \ref{sec:Experimental-setup}. In Section \ref{sec:evaluation}, we discuss experimental results and finally, make the conclusion and propose future works in Section \ref{sec:conclusion}.\\

\section{Related Works}
\label{sec:related-works}
Different techniques are proposed based on physiological and behavioral clues used for identification. Four types of Spoof attacks mostly threaten ASVs. 
Among them, replay attacks focus on using a pre-recorded speech from a genuine speaker in order to deceive ASVs. Speech could be recorded without the speaker’s consent or even it can be generated by concatenating different parts of speech from the genuine speaker. High tech devices including smartphones, laptops and recorders, designed for recording, complicate the situation, even more. As a result, it can be the most potent threat to ASVs. The vulnerability of ASVs to replay attacks first was investigated by \cite{Lindberg1999}, where authors reported a significant increase in Equal Error Rate (EER) and False Acceptance Rate (FAR) for ASVs. Other studies \cite{Villalba2010}\cite{Villalba2011} demonstrated that using Joint Factor Analysis (JFA) in which FAR increased for the attack in this category. In \cite{Ji2017}, for a text-independent system with GMM-UBM classifier considerable increase in EER for ASVs is reported. As mentioned, there are various low-cost ASVs that have saturated the market from one hand, and on the other hand, easiness and availability of devices required for replay attacks, lead to proposing countermeasures for both sides \cite{Lee2013}. Furthermore, for a replay attack, there is no need for special speech processing techniques \cite{Wu2014} and unlike other types of spoofing, they can be used with less knowledge about ASVs. Hence, they are more engaged in spoofing attacks.  

Therefore, in this study, considering the importance of replay spoofing attacks, we focus on this type of attack, and we will examine countermeasures for this threat. Among one of the first countermeasures for a replay attack, \cite{Shang2010} employed fixed pass-phrase on a text-dependent ASVs. The detector stores past access attempts with new attempts and then make the decision. Results demonstrated improvement in EER while they detected and recognized playback utterances. In \cite{kinnunen2012vulnerability}, authors use spectral ratio and modulation indexes in order to detect spoofing in far-field recordings where noises increase in recorded speech due to the distance. As a result of the long-distance, the spectrum turns into a flattened one and then the ability for modulation index is reduced. For classification purposes, an SVM approach was used. Their results show that the FAR of text-independent JFA is reduced for the ASVs. Wang \textit{et al.} \cite{Wang2011} claims that licit recordings have just a specific type of channel noise which is mixed with additional noise of environment when a replay attack is presented in far-field. A GMM-UBM classifier was employed in order to detect the replay attack and the results show a significant decrease in EER.

%In spite of the mentioned replay anti-spoofing approaches, the most serious works in this area were presented after the publication of BTAS 2016 competition \cite{BTAS} and ASVspoof Challenge 2017 dataset \cite{Kinnunen2017} and continued with the introduction of the ASVspoof Challenge 2019 \cite{Todisco2019}.

%\subsection{Replay Spoofing Attacks Related Works}
Ji \textit{et al.} \cite{Ji2017}, captured CQCC features and feed them to a decision tree classifier, Results showed EER of 10.8\% for this study on ASVspoof 2017 dataset. Adiban \textit{et al.} \cite{Adiban2017} combined different features including Mel-Frequency Cepstral Coefficients (MFCC), RASTA-PLP, Modified Group Delay, CQCC, i-vector, and Linear Prediction Cepstral Coefficients (LPCC) and fed them to different classifiers such as SVM, Multi-Layer Perceptron (MLP) neural networks and GMMs on ASVspoof 2017 dataset. The best performance was 10.31\% for EER. Shim \textit{et al.} \cite{Shim2018} investigated the problem of replay spoofing attack detection and noise classification using multi-task learning on playback devices, recording environments and devices. Results showed a 30\% improvement from 13.57\% to 9.56\% in terms of EER on ASVspoof 2017 dataset. A Discrete Fourier Transform (DFT) was used in \cite{Alam2018} for ASVspoof 2017 dataset. Then, a normalization was applied to features in the q-log domain. For feature reduction,  Principal Component Analysis (PCA) performed on initial feature. The best performance was at the EER of 11.19\%. Anti-spoofing task on ASVspoof 2017 dataset was performed in \cite{Lavrentyeva2017} where a CNN used as a deep learning approach with Max-Feature-Map activation function. The proposed approach yielded 6.73\% as EER. Long-term temporal envelopes also extracted from sub-band signals using Frequency Domain Linear Prediction (FDLP) for feature extraction and GMM and CNN used for classification were presented in \cite{Wickramasinghe2018}. The reported EER on ASVspoof 2017 dataset was 9.70\%. Sailor \textit{et al.} \cite{Sailor2018} reported EER of 8.89\%  on ASVspoof 2017 dataset using ConvRBM-CC as features and GMMs as classification. 

In the work presented in \cite{cai2019dku}, inverted Mel-frequency cepstral Coefficients (IMFCC), short time Fourier transform (STFT), group delay gram (GD gram) and joint gram are used as features. Extracted features fed into a residual neural network (ResNet) classifier. The best results obtained in score level fusion on the ASVspoof 2019 dataset, which are 0.66\% and 0.0168 in terms of EER and t-DCF, respectively. Li \textit{et al.} in \cite{li2019anti}, implemented on ASVspoof 2019 dataset utilizes multiple spectral features within the network, such as MFCC, CQCC, Fbank, etc., and also butterfly unit (BU) for multi-task learning. Results showed 0.67\% and 0.0148 in terms of EER and t-DCF, respectively. In \cite{alzantot2019deep}, various features, including MFCC, CQCC, and STFT are extracted and then fed into ResNet classifier. Finally, in the fusion level EER of 0.28\% and t-DCF of 0.0074 are achieved. The study investigated in \cite{das2019long}, proposed a method for replay spoofing detection based on various long range acoustic features and Deep Neural Networks as classifiers. Their best combined system obtained t-DCF of 0.1381  and EER of 5.95\% for physical access on AVSpoof 2019 dataset. The study in \cite{lavrentyeva2019stc} is another work in which the proposed system is implemented using Light CNN (LCNN) based on various acoustic features such as CQT, LFCC, and Discrete Cosine Transform (DCT). The best result reported in this study is obtained in score level fusion which is 0.54 and 0.0122 in terms of ERR and t-DCF, respectively. In \cite{chettri2019ensemble}, authors utilize different spectral features, including delta and acceleration MFCC (SDA MFCC), Inverted MFCC (IMFCC), CQCC, sub-band centroid magnitude coefficients (SCMC) features and i-vector in order to detect spoofing using various shallow and deep classifiers such as GMM, SVM, CNN, Convolutional Recurrent Neural Network (CRNN), 1D-CNN and Wave U net, resulting in EER of 5.43 and t-DCF of 0.1465.

Table \ref{tbl:related-works} summaries the replay attack countermeasures investigated in this section.

\begin{figure*}[ht]
\centering
\includegraphics[width=0.80\textwidth]{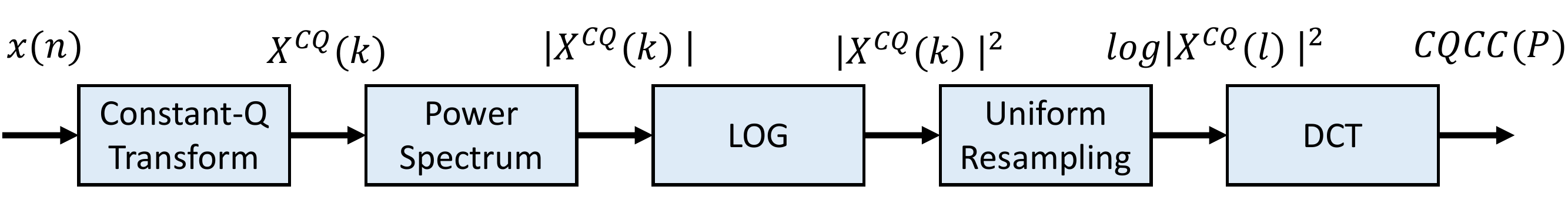}
%\vspace{-0.4cm}
\caption{Block diagram of CQCC feature extraction \cite{Todisco2016}.}
%\vspace{-0.2cm}
\label{fig:CQCC}
\end{figure*}

\begin{table*}[h!]
\begin{center}
\caption {Summary of the replay attack Countermeasure (CM) Methods, and Results.}
\label{tbl:related-works}
\begin{tabular}{| c | c  | c |}\hline
Study & CM Method & EER\% / t-DCF  \\ \hline

{\color{blue}Z. Ji, \textit{et al.}} \cite{Ji2017}& CQCC + GMM mean supervector-Gradient Boosting & 10.8 / - \\ \hline
{\color{blue}M. Adiban, \textit{et al.}} \cite{Adiban2017}& MFCC, RASTA-PLP, CQCC, LPCC and i-vector + GMM, MLP and SVM & 10.31 / - \\ \hline
{\color{blue}H.-J. Shim \textit{et al.}} \cite{Shim2018}& Noise detection + neural network and Multi-task learning & 9.56-13.57 / - \\ \hline
{\color{blue}M. J. Alam, \textit{et al.}} \cite{Alam2018}& DFT-based features + feature normalization + PCA & 11.9 / - \\ \hline
{\color{blue}G. Lavrentyeva, \textit{et al.}} \cite{Lavrentyeva2017}& Max-Feature-Map activation + CNN & 6.37 / - \\ \hline
{\color{blue}B. Wickramasinghe, \textit{et al.}} \cite{Wickramasinghe2018}& FDLP (TC and RC) + GMM and CNN & 9.70 / -\\ \hline
{\color{blue}H. B. Sailor, \textit{et al.}} \cite{Sailor2018}& ConvRBM-CC + GMM & 8.89 / - \\ \hline

{\color{blue}W. Cai, \textit{et al.}} \cite{cai2019dku}& IMFCC,STFT, GD gram, joint Gram + ResNet & 0.66 / 0.0168 \\ \hline
{\color{blue}R. Li, \textit{et al.}} \cite{li2019anti}& MFCC,CQCC, Fbank + BU & 0.67 / 0.0148 \\ \hline
{\color{blue}M. Alzantot, \textit{et al.}} \cite{alzantot2019deep}& MFCC,CQCC, STFT + ResNet & 0.28 / 0.0074 \\ \hline
{\color{blue}R. K. Das, \textit{et al.}} \cite{das2019long}&  long  rangeacoustic  features + DNN & 5.95 / 0.1381 \\ \hline

{\color{blue}G. Lavrentyeva, \textit{et al.}} \cite{lavrentyeva2019stc}&    CQT,  LFCC  and  DCT + LCNN & 0.54 / 0.0122 \\ \hline

{\color{blue}B. Chettri, \textit{et al.}} \cite{chettri2019ensemble}&    Spectral features, i-vector + Deep and Shallow Classifiers & 5.43 / 0.1465 \\ \hline

\end{tabular}
\vspace{-6mm}
\end{center}
\end{table*}

\section{Models}
\label{sec:models}
\subsection{Constant Q Cepstral Coefficient}
\begin{figure}[H]
\centering
\includegraphics[width=0.50\textwidth]{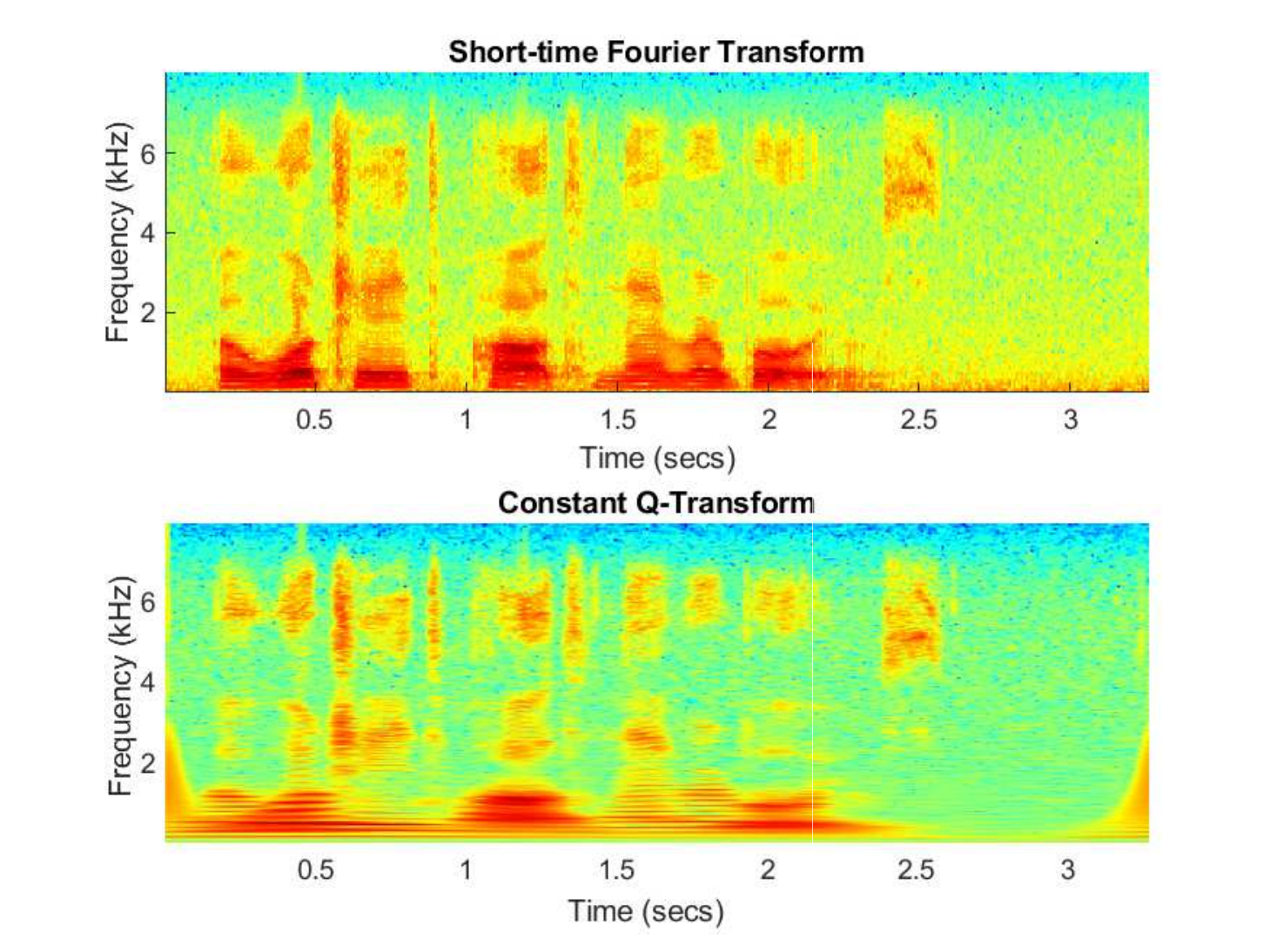}
%\vspace{-0.4cm}
\caption{A comparison of CQT and STFT spectrogram.}
%\vspace{-0.2cm}
\label{fig:CQT-STFT}
\end{figure}

In this work we used CQCCs. These features utilize Constant Q Transform (CQT) \cite{Brown1991} which is originally presented for tasks related to music processing \cite{Schorkhuber2010} and later it was employed in speaker verification tasks \cite{Todisco2016}. Despite many cepstral based features utilizing prevalent Fourier Transform (FT), CQCCs employ Constant Q Transform (CQT) which uses geometrically spaced frequency bins \cite{Todisco2016}. The main difference between FT and CQT is that FT uses fixed time-frequency resolution, however, CQT utilizes a constant Q factor which leads to an increase in frequency resolution at lower frequencies and also a better time resolution at higher frequencies \cite{Todisco2016a}. The Q factor is the ratio between the center frequency $f_k$ and the bandwidth $\delta_f$ defined as.
\begin{equation}
    Q = \frac{f_k}{\delta_f}.
\end{equation}
Details about CQT can be found in \cite{Todisco2016a}. The CQCC extraction steps are introduced as follows: Initially, for the given signal $x(n)$, CQT is engaged to obtain the spectrum. Then, the power spectrum and subsequently the logarithmic power spectrum are calculated. Eventually, in order to attain the cepstrum of $x(n)$, we apply the DCT as the inverse transformation in the spectrum logarithm. The extraction steps of CQCC are shown in Fig. \ref{fig:CQCC}. In addition, a comparison of the CQT and STFT spectrogram for a speech signal is illustrated in Fig. \ref{fig:CQT-STFT}. 

\subsection{Autoncoders}
\label{sec:AutoEncoders}
\begin{figure}[H]
\centering
\includegraphics[width=0.45\textwidth]{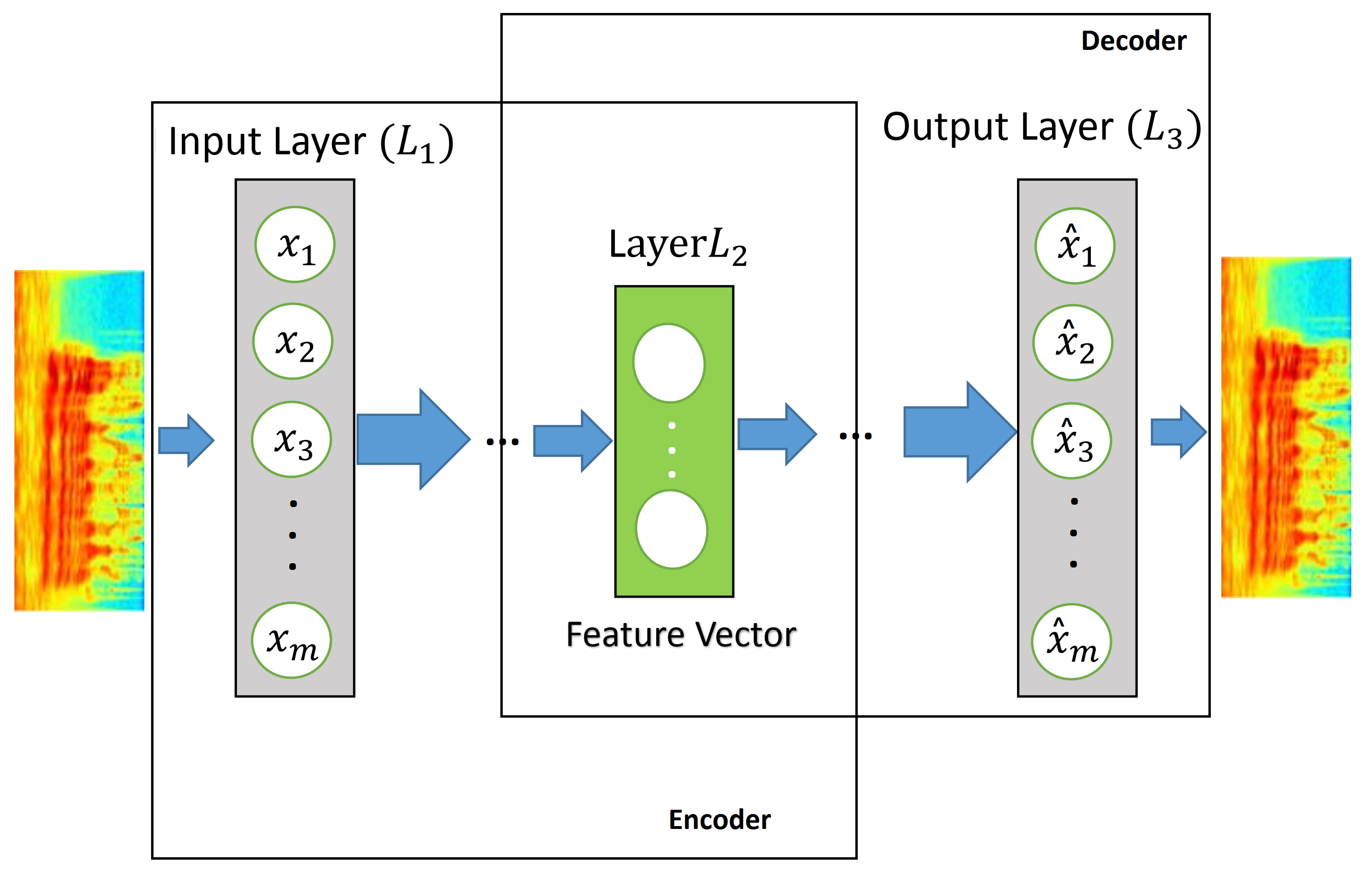}
%\vspace{-0.4cm}
\caption{Diagram of the Autoencoder.}
%\vspace{-0.2cm}
\label{fig:AE}
\end{figure}

Autoencoder (AE) is a generative model that can be used to extract substantial information and features. Accordingly, an autoencoder is a neural network devised for the purpose of learning an identity function in an unsupervised setting to reconstruct high dimensional input data \{$x_1; x_2; ...; x_m$\} as outputs \{$\hat{x}_1; \hat{x}_2; ...; \hat{x}_m$\}. As mentioned in Sec. \ref{sec:introduction}, AE is capable of capturing latent variables hidden in input samples. Furthermore, AE extracts most informative and independent features to compress the significant information of input samples \cite{hsu2017unsupervised}. Considering the sparse nature of noise, AE can both capture the sparse features of the noisy input and also to map them to a features space, capable of discriminating noisy and clean samples. Accordingly, classifier is able to discriminate both type of samples with high accuracy. The structure of the autoencoder is depicted in Fig. \ref{fig:AE}. According to this structure, the input data feature is reduced to reduced compact ones, which can be considered as a feature vector. As shown in Fig. \ref{fig:AE},  $L_2$ is a hidden layer. Eq. \ref{eq:AE} defines the activation of unit $i$ in layer $l$.
\begin{equation}
\label{eq:AE}
    a_{i}^{(l)} = f(\sum_{j=1}^{n}W_{ij}^{(l-1)}a_j^{(l-1)} + b_j^{(1)}),
\end{equation}
where $W$ represents weight and $b$ denotes bias parameters. According to Fig. \ref{fig:AE}, the input layer is considered as $a^{(1)}=x$, and in the output layer, $a^{(3)}= \hat{x}$ . In addition, in the hidden layers, the sigmoid function is used as the activation function $f$. However, the linear function is engaged in the output layer. The objective function will be defined as:
\begin{equation}
\label{eq:AE-objective}
    J(W,b) = \frac{1}{m}\sum_{i=1}^{m}(\frac{1}{2}||x_i-\hat{x}_i||^2) + \frac{\lambda}{2}\sum_{l=1}^{n_1 - 1}\sum_{i=1}^{s_1 - 1}\sum_{j=1}^{s_l + 1} (W_{ji}^{(l)})^2,
\end{equation}
where the parameters $\lambda$, $n_l$ and $s_l$  represent the strength of regularization, the number of layers in the network and the number of units in layer $L_1$, respectively. In the training phase, the objective function will be minimized with respect to the parameters of $W$ and $b$.

\subsection{Siamese Networks}
\label{sec:Siamese-Network}
\begin{figure*}%htbp]
\centering
\includegraphics[width=0.75\textwidth]{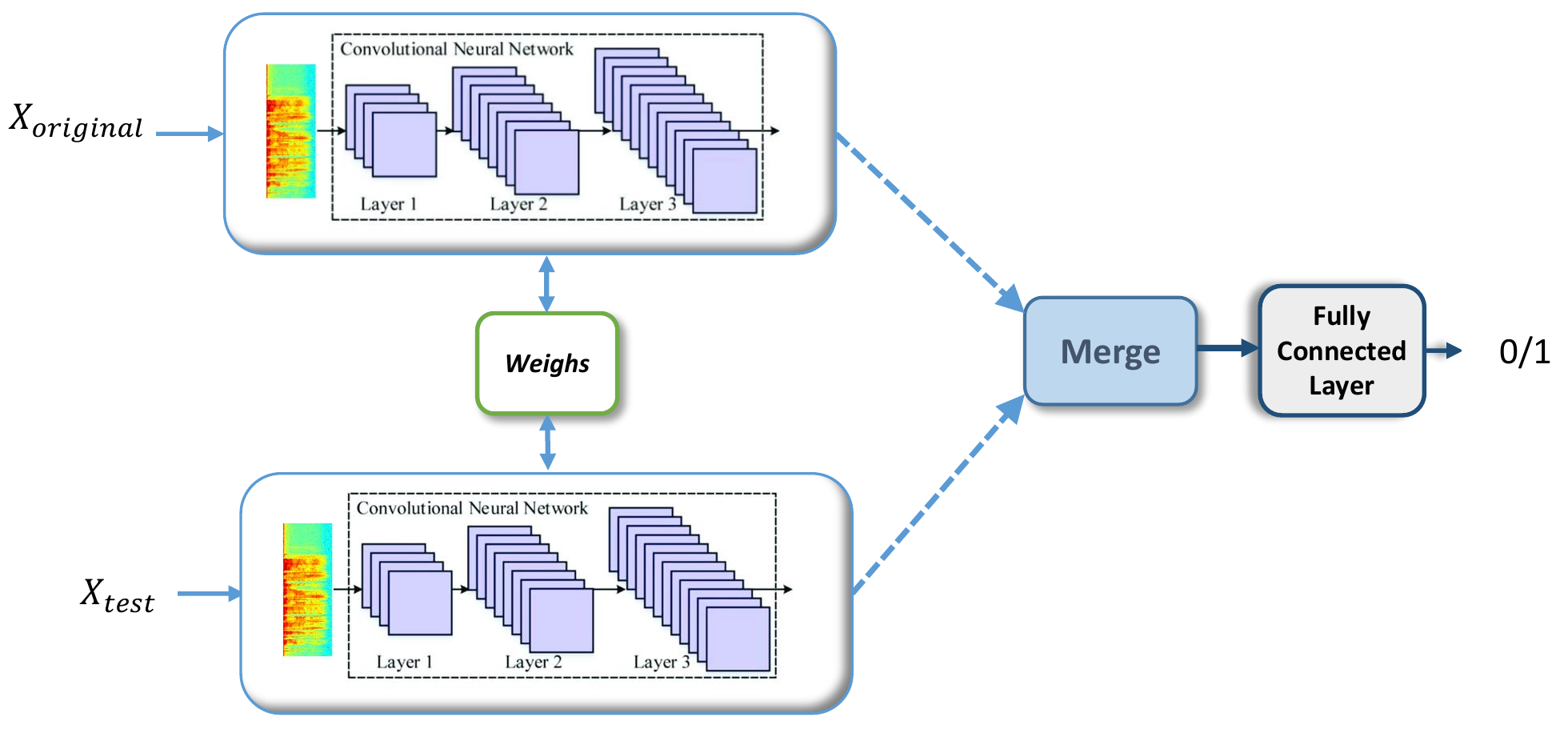}
%\vspace{-0.4cm}
\caption{Architecture of Siamese Networks.}
%\vspace{-0.2cm}
\label{fig:Siamese}
\end{figure*}

Siamese Networks are one among all types of neural networks first used by \cite{ChopraS.HadsellR.2005}\cite{BromleyJ.BentzJ.W.ButtouL.GuyonI.LeCunY.MooreC.SackingerE.Shah1993} for different purposes. Most of the works focused on their application in verification tasks such as face recognition, signature recognition, etc. Siamese networks take the sample as input and then maps it onto a new latent space where similar samples have shorter distance than non-similar ones. So the whole idea is finding a “Target Space” where the semantic distance between inputs is found. Thus, the Siamese networks can be very useful where the training data does not contain sufficient information needed for classification.\\
Hence, it can be realized that Siamese networks can be applied in verification tasks. Spoofing detection is also a matter of verification process of whether a voice is genuine or not. In this work, considering the fake nature of spoof, a similarity measure can be engaged to compare features extracted from both data given to Siamese and then use another neural layer for the final decision. 

Architecture for this network is given in Fig. \ref{fig:Siamese}. It includes two identical deep neural networks with the same configurations in terms of weights, hyper-parameters, etc. Each one of these neural networks is called “leg” of the network and both are identical. One network is trained by input samples and then a copy of the network is used with the same weights as the other leg. The network can be a simple Multi-Layer Perceptron (MLP) or other types of deep neural networks such as CNN or RNN. The results are combined mostly using a combination function and then the output is fed to a fully connected network that is trained to produce a metric to show if two samples are the same or different based on output given by the combination function. So the task here is to minimize the loss function based on the combination function. The final fully connected layer can be performed by a sigmoid function or one single neuron.

\section{Proposed System}
\label{sec:proposed-system}
\begin{figure*}[htbp]
\centering
\includegraphics[width=0.95\textwidth]{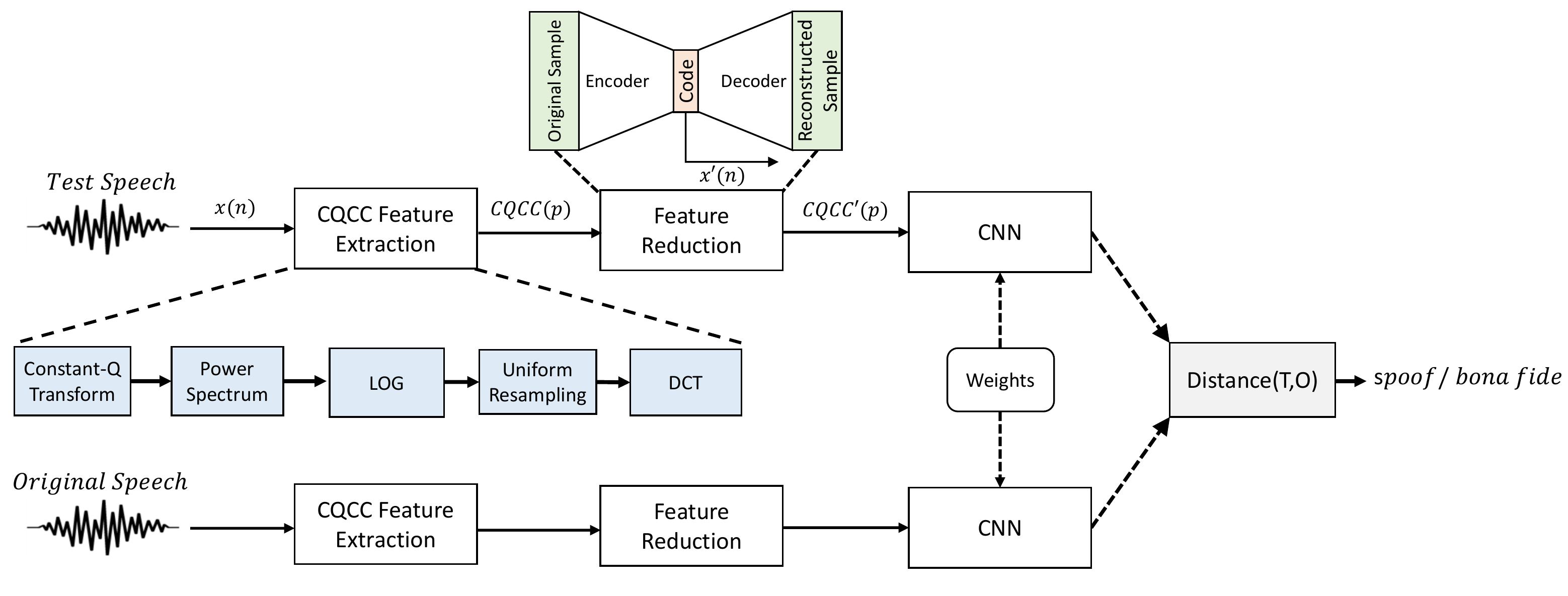}
%\vspace{-0.4cm}
\caption{Diagram of our proposed Siamese network.}
%\vspace{-0.2cm}
\label{fig:framework}
\end{figure*}
In this section, we will give information about the Model we employed for each step of the whole framework. The proposed system is depicted in Fig. \ref{fig:framework}. For the first step, CQCC features with different dimensions are extracted from the initial voice using the CQCC feature extractor. Thereupon, for encoding features, we used an $X\times Y\times X$ structure where $X$ is the dimension size of the extracted features and $Y$ is the bottleneck dimension size. For training, we engaged a fine-tuning back propagation, so the parameters will be improved in each iteration. Then we used a Convolution Neural Network (CNN) similar to the one used in \cite{Yang2014LearnCN}. CNN consisted of 3 convolutional layers, 3 max-pooling layers and, 2 Fully Connected (FC) layers. Max-pooling operator have shown to be sensitive to noisy input which is an important features. Noisy input is one of the important characteristics of a replay attack \cite{ma2018deep}. Two independent convolutional networks were used in every layer with Max-Feature-Map (MFM) \cite{wu2015lightened} as the activation function. For compression purposes, a max-pooling kernel was used with a size of 2$\times$2. For the FC layer, we used a softmax function to discriminate between spoof and genuine samples. For improving the performance, we employed  a “Highway” structure from \cite{Greff2015} in the final layer which regulates information transfer considering this fact that each neural network is a highway, where unimpeded information can flow over several layers without attenuation. We used ReLU as an affine transform function and sigmoid as an activation function:
\begin{equation}
\label{eq:highway}
    y = H(x,W_H)*T(x,W_T) + x.(1 - T(x,W_T))
\end{equation}
In Eq. \ref{eq:highway}, $H,T$ are the affine transform function and activation function, respectively. The weighting matrices $W_h, W_T$  are updated in training process. $x$ represents the input and the dot operator (.) is used to denote element-wise multiplication. Finally $y$ is output of highway function. 
Probability of a voice to be a spoof or not is calculated by a softmax layer using the following equation:
\begin{equation}
\label{eq:softmax}
    P(c_i|z) = \frac{e^{z^Tw}}{\sum_{i=1}^{C}e^{z_i^Tw}},
\end{equation}
where $c_i$ is the $i^{th}$ class that we try to calculate whether sample $x$ is a member of this class or not. $z$ is the given feature vector and $C$ is the number of whole classes, which in our case is two, spoof and genuine. 

One of the trained CNNs is fed with genuine fixed-length vectors obtained from autoencoder and the other ones are fed with the test sample. If they are similar they will get the same class and otherwise, they will be placed in different classes. As \cite{Computing2017}, we tried to use dropout for preventing overfitting with mini-batches of 200 samples.
For loss function we used Cross Entropy:
\begin{equation}
\label{eq:CE}
    CE = - \sum_{i=1}^{C'}t_i\log(f(s_i))-(1-t_i)\log(1-f(s_i)),
\end{equation}
where $t_i$ is the true label of class $i$, $f(s_i)$ is the calculated label for sample $s_i$ and $C'$ indicates the number of total classes in the dataset which is 2 in our case. Finally, a simple loss function is used to calculate the distance between each feature to take the value of 1 (spoof) if it is more than 0.5 and 0 (genuine) otherwise. 

\section{Experimental Setup}
\label{sec:Experimental-setup}
\subsection{Dataset}
The ASVspoof 2019 challenge \cite{Todisco2019} is introduced to expand the goals of the previous challenges ASVspoof 2015 \cite{Wu2015} and ASVspoof 2017 \cite{Kinnunen2017}. The main purpose in ASVspoof 2015 was to introduce countermeasure systems for detecting spoofed/non-spoofed speech in which spoofed speech was implemented based on either text-to-speech (TTS) or voice conversion (VC) approaches. In addition, the 2017 ASVspoof Challenge was presented to guide studies to introduce countermeasure systems for detecting replay spoofing attacks. Subsequently, the ASVspoof 2019 challenge was presented to complete the objectives of the two previous challenges and provided two subsets including Logical Access (LA) and Physical Access (PA). In the PA scenario, the spoofing attacks are based on replay attacks where an adversary tries to record a genuine speech and replay it in order to deceive the ASV system. PA includes three subsets: Training, Development, and Evaluation sets.\\
For training and development, a dataset was created using a combination of three room sizes, with three levels of reverberation and three different speaker-to-ASV microphone distances, and in total 27 configurations. The replay attack dataset comprises nine different configurations from three different categories of distances and three different qualities.  Evaluation dataset consisted of 137457 trials including both replay spoofed speech and bona fide with different configurations with unique speakers. The statistics of each of subsets is summarized in Table \ref{tbl:database-2019}.

\begin{table}[ht]
\begin{center}
\caption {Physical Access Scenario Statistics for ASVSPOOF 2019 Database}
\label{tbl:database-2019}
\begin{tabular}{| l | c  | c | c | c |}\hline
\multirow{2}{*}{Subset}&\multicolumn{2}{c|}{\#Speaker}&\multicolumn{2}{c|}{\#Utterance}\\
{}&\multicolumn{1}{c}{Male}&{Female}&\multicolumn{1}{c}{Bona fide}&{Spoof}\\ \hline
{Training}&{8}&{12}&{5400}&{48600}\\
{Development}&{4}&{6}&{5400}&{24300}\\
{Evaluation}&{21}&{27}&{18090}&{116640}\\ \hline
{Total}&{33}&{45}&{28890}&{189540}\\ \hline
\end{tabular}
\vspace{-6mm}
\end{center}
\end{table}

\subsection{Evaluation Metrics}
\label{sec:eval-met}

The ASVspoof 2019 challenge is a binary classification task, in which utterances from real humans are labeled as a positive class and spoof attacks are labeled as a negative class. Tandem Detection Cost Function (t-DCF) \cite{Kinnunen2018} is adopted by ASVspoof 2019 as a standard measure metric, which is based on detection theory and can be specified for the envisioned application. Isolation of different systems (such as CM and ASV) is a key feature of tandem systems and t-DCF. We also use EER as an additional metric to measure our system performance.

\section{Experimental Evaluation}
\label{sec:evaluation}
%In this section, we will present a discussion about the experimental results. Firstly, the baseline system will be introduced in the subsection 5.1, in the following the main results of our proposed system with different configurations will be presented in the subsection 5.2 and finally, in the subsection 5.3 our results are compared with baseline system and then the analysis is performed different aspects of innovation in our work.
\subsection{Baseline System}
\label{sec:baseline}
The organizers of the challenge ASVspoof 2019 have introduced two systems as a baseline \cite{Todisco2019} for participators. The proposed method in the baseline system is based on CQCC and  Linear Frequency Cepstral  Coefficients (LFCC) \cite{Sahidullah2015} features and GMM classifiers. Accordingly, they extract CQCC and LFCC features from training data and then two GMMs (one GMM for the bona fide and the other for spoofed data) with 512 components to learn the model by EM iterations (training GMMs). In the next step, the score for each trial is computed using these GMMs. The results for the development set and evaluation set of physical access scenario in terms of t-DCF and EER are presented in Table \ref{tbl:baseline-results}.

\begin{table}[ht]
\begin{center}
\caption {t-DCF and EER results for two baseline countermeasures on physical access scenarios for both development set and evaluation set \cite{Todisco2019}\cite{Kinnunen2016}.}
\label{tbl:baseline-results}
\begin{tabular}{| c | c  | c | c | c |}\hline
\multirow{2}{*}{Baseline System}&\multicolumn{2}{c|}{Development Set}&\multicolumn{2}{c|}{Evaluation Set}\\ \cline{2-5}
{}&\multicolumn{1}{c|}{EER\%}&{t-DCF}&\multicolumn{1}{c|}{EER\%}&{t-DCF}\\ \hline\hline
{LFCC-GMM}&{11.96}&{0.2554}&{13.54}&{0.3017}\\ \hline
{CQCC-GMM}&{9.87}&{0.1953}&{11.04}&{0.2454}\\ \hline
\end{tabular}
\vspace{-6mm}
\end{center}
\end{table}

\subsection{Proposed System Configuration}
As we have mentioned, our proposed system consists of three main parts. First, extracting the CQCC features with different dimensions from input speech. Second, dimensionality feature reduction using autoencoder to different dimensions. Finally, using the Siamese network as a classifier. This process includes the Training and Evaluation phases:
\subsubsection{Training Phase}
To train our system, we extract CQCC features from training data (for both bona fide and spoofed data). In the following, we trained an autoencoder with different bottleneck dimensions using the extracted features separately with Siamese networks. Accordingly, we take into account two objectives of training the autoencoder. First, we try to reduce the data dispersion. More importantly, autoencoder is trained by considering the noise in spoofed speech, leading to achieving more valuable data for spoofing detection. As the final step, we trained Siamese Networks (including two CNNs) as our classifier. In order to train these CNNs, first, we divided our dataset into two balanced parts (one part is used for first CNN and another part for the second one). Afterwards,  we randomly selected data form each part without replacing it and then apply each selected data considering their labels to the network assigned to that part. Finally, CNNs learn whether two input data are from equal classes or not and subsequently their shared weights are updated. Here, different configurations of CNNs, indicated as “Siamese Network Config", were used to observe the impacts of using various filters with different sizes on performance. Considering the effects of using different feature extraction methods, we used three different configurations of CNN. In this regard, we run our systems on development data to obtain our configurations and tune their parameters.  Therefore, the first configuration, indicated by Config. 1, contains three layers of convolutional and average-pooling layers and 1 hidden layer. It holds 160, 200, and 100 filters, respectively. In the hidden layer, we have 300 nodes and 3 pooling width. For the second configuration (Config. 2), we have the same architecture, with a change in the number of the hidden nodes to 500. In the third configuration (Config. 3), we just replaced average-pooling with max-pooling to observe the impact of the max filter instead of average-pooling one.

\subsubsection{Evaluation Phase}
In the evaluation phase, similar to the training phase, we extract different dimensions of the CQCC features from the evaluation set. We then used an autoencoder to reduce the dimensions of the extracted features. At the classification level, we applied evaluation data to one of the trained CNN in the previous step. And the input of another CNN is a fixed bona fide speech which is randomly selected from the training set. So for each evaluation data, we will check whether the input data matches the fixed bona fide speech. If it matches, we will consider it as a bona fide, otherwise, it will be considered as a spoofed speech. Finally, we repeat this cycle with 100 fixed bona fide data which are randomly selected from the training set without considering their gender or number of speakers and finally, we get the majority votes. As mentioned in the training phase, it should be reminded that we used development trials in order to tune the parameters of our system. 
\subsection{Siamese Networks Results and Analysis}
\label{sec:siamese-analysis}
Results obtained from different classifier setups are investigated in this section. Table \ref{tbl:proposed-results} shows EER and t-DCF achieved from different dimensions of CQCC and different configurations of our classifier without any dimensionality reduction on the development set and evaluation set. As shown in Table \ref{tbl:proposed-results}, the best results in each classifier configuration belong to 90-dimensional CQCC. Among these results, the best result is achieved from Siamese Config. 3. It can be seen this result improves the baseline system 10.42\% and 0.2344 in terms of EER and t-DCF, respectively.
\\
\textbf{Discussion:} As explained in the introduction, using descriptive CQCC through a higher dimension of features can be effective in the spoofing detection task due to the fact that it gives a better resolution at different frequencies. This leads to a detailed description of noisy frequencies and usual ones. On the other hand, increasing the dimension of the features might trigger redundancy. Thus, we are looking for a balance between higher resolution and lower redundancy. Experimental results show that 90-dimensional CQCC achieves the equilibrium point. In addition, it is obvious that results for Config. 3 are better than the other two configurations. The key point for this configuration is that it has fewer hidden nodes than the second configuration and so we can realize that increasing the number of hidden nodes is helpful up to some point and after that the performance decays. Furthermore, it shows that using max-pooling layer can be more helpful than average-pooling. This also comes back to the noisy nature of the given task, where the noises’ effects can be removed using the average-pooling \cite{ma2018deep}, while max-pooling will result in selecting noisier values at the selection layer. 
\begin{table}[htbp]
\begin{center}
\caption {CQCC dimension effects on the performance of the proposed system with different configurations of CNN configurations noted as "Siamese Configs".}
\label{tbl:proposed-results}
\begin{tabular}{|p{0.2cm}p{0.6cm}|c|c|c|c|c|c|}\cline{3-8}
\multicolumn{1}{c}{}&\multicolumn{1}{c|}{} & \multicolumn{6}{c|}{Siamese Network Configs.} \\ \cline{3-8}
\multicolumn{1}{c}{}&\multicolumn{1}{c|}{} & \multicolumn{2}{c|}{Config. 1} & \multicolumn{2}{c|}{Config. 2} & \multicolumn{2}{c|}{Config. 3} \\ \hline
\multicolumn{1}{|c|}{Set}&CQCC Dims.& \multicolumn{1}{c|}{EER\%} & \multicolumn{1}{c|}{t-DCF} & EER\% & t-DCF & EER\% & t-DCF \\ \hline\hline 
\multirow{4}{*}{\rotatebox[origin=c]{90}{Dev.}}&\multicolumn{1}{|c|}{30} & 2.17 & 0.0411 & 2.35 & 0.0437 & 1.96 & 0.0391 \\ \cline{2-8}
&\multicolumn{1}{|c|}{60} & 1.29 & 0.0240 & 1.84 & 0.0319 & 1.65 & 0.0324 \\ \cline{2-8}
&\multicolumn{1}{|c|}{90} & \textbf{1.09} & \textbf{0.0223} & \textbf{1.38} & \textbf{0.0245} & \textbf{0.92} & \textbf{0.0183} \\ \cline{2-8}
&\multicolumn{1}{|c|}{120} & 1.46 & 0.0247 & 1.93 & 0.0386 & 1.24 & 0.0207 \\ \hline\hline

\multirow{4}{*}{\rotatebox[origin=c]{90}{Eval.}}&\multicolumn{1}{|c|}{30} & 6.80 & 0.1550 & 7.12 & 0.1649 & 5.66 & 0.1371 \\ \cline{2-8}
&\multicolumn{1}{|c|}{60} & 4.19 & 0.1072 & 5.57 & 0.1323 & 3.78 & 0.0883 \\ \cline{2-8}
&\multicolumn{1}{|c|}{90} & \textbf{3.27} & \textbf{0.0745} & \textbf{4.13} & \textbf{0.1086} & \textbf{3.02} & \textbf{0.0627} \\ \cline{2-8}
&\multicolumn{1}{|c|}{120} & 4.73 & 0.1251 & 6.29 & 0.1528 & 4.11 & 0.0854 \\ \hline

%Siamese Config1 &\multicolumn{1}{c}{} & \multicolumn{1}{c|}{} & Siamese Config2 & \multicolumn{1}{c}{} & \multicolumn{1}{c|}{} & Siamese Config3& \multicolumn{1}{c}{} & \multicolumn{1}{c|}{} \\ \cline{1-1}\cline{4-4}\cline{7-7}
%CQCC Dimensions &\multicolumn{1}{c}{EER\%} & \multicolumn{1}{c|}{t-DCF} & CQCC Dimensions &\multicolumn{1}{c}{EER\%} & \multicolumn{1}{c|}{t-DCF} & CQCC Dimensions &\multicolumn{1}{c}{EER\%} & \multicolumn{1}{c|}{t-DCF} \\ \hline\hline
%30 & 6.80 & 0.1550 & 30 & 7.12 & 0.1649 & 30 & 5.66 & .1371 \\ \hline
%60 & 4.19 & 0.1072 & 60 & 5.57 & 0.1323 & 60 & 3.78 & 0.0883 \\ \hline
%90 & \textbf{3.27} & \textbf{0.0745} & 90 & \textbf{4.13} & \textbf{0.1086} & 90 & \textbf{3.02} & \textbf{0.0627} \\ \hline
\end{tabular}
\vspace{-6mm}
\end{center}
\end{table}

\subsection{Autoencoder Performance}
\begin{figure}[htbp]
\centering
\includegraphics[width=0.45\textwidth]{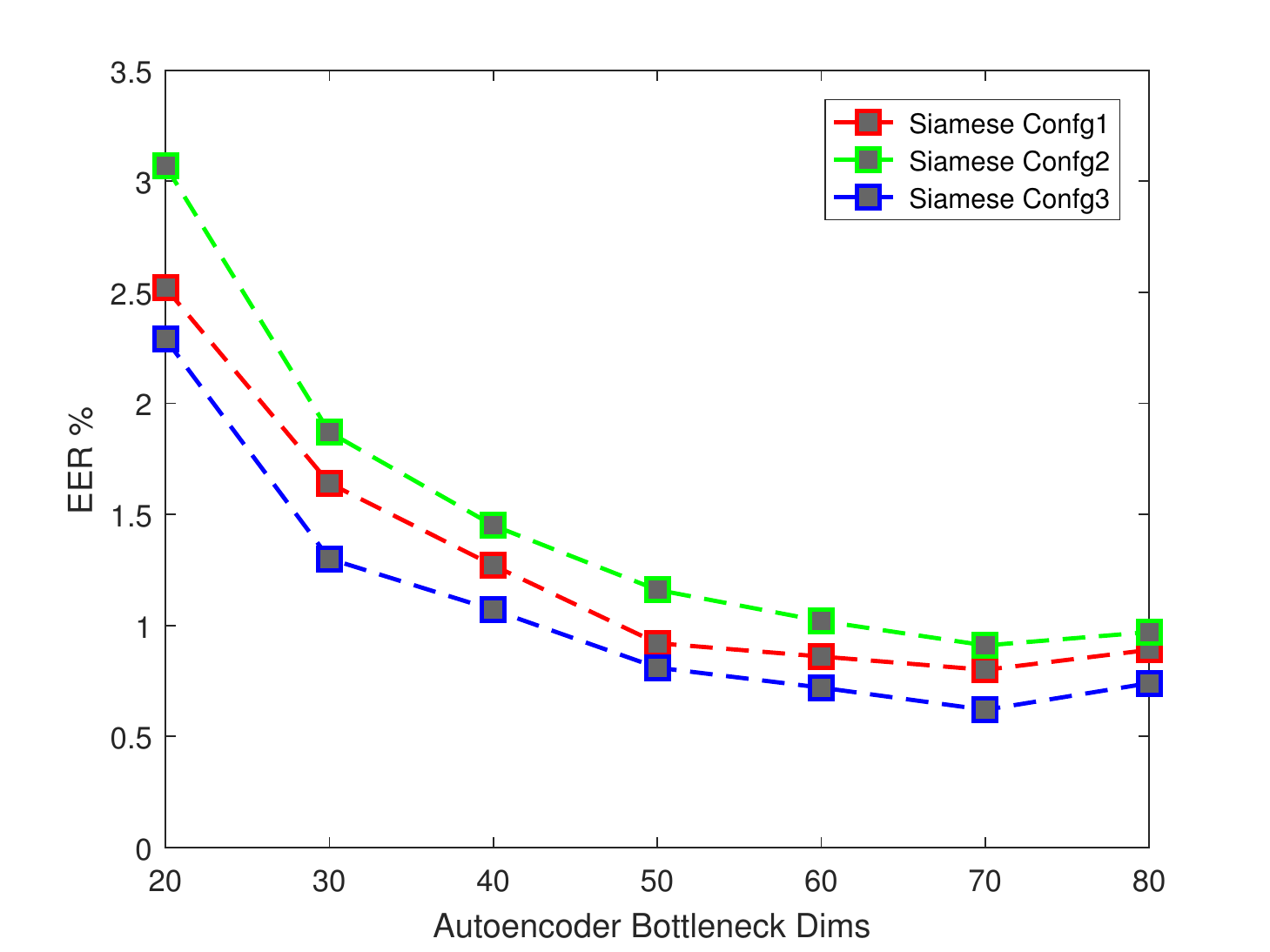}
%\vspace{-0.4cm}
\caption{Effects of dimension reduction by autoencoder on EER for evaluation set with CQCC vector size of 90 on the proposed system.}
%\vspace{-0.2cm}
\label{fig:CQCC-AE-EER}
\end{figure}

\begin{figure}[htbp]
\centering
\includegraphics[width=0.45\textwidth]{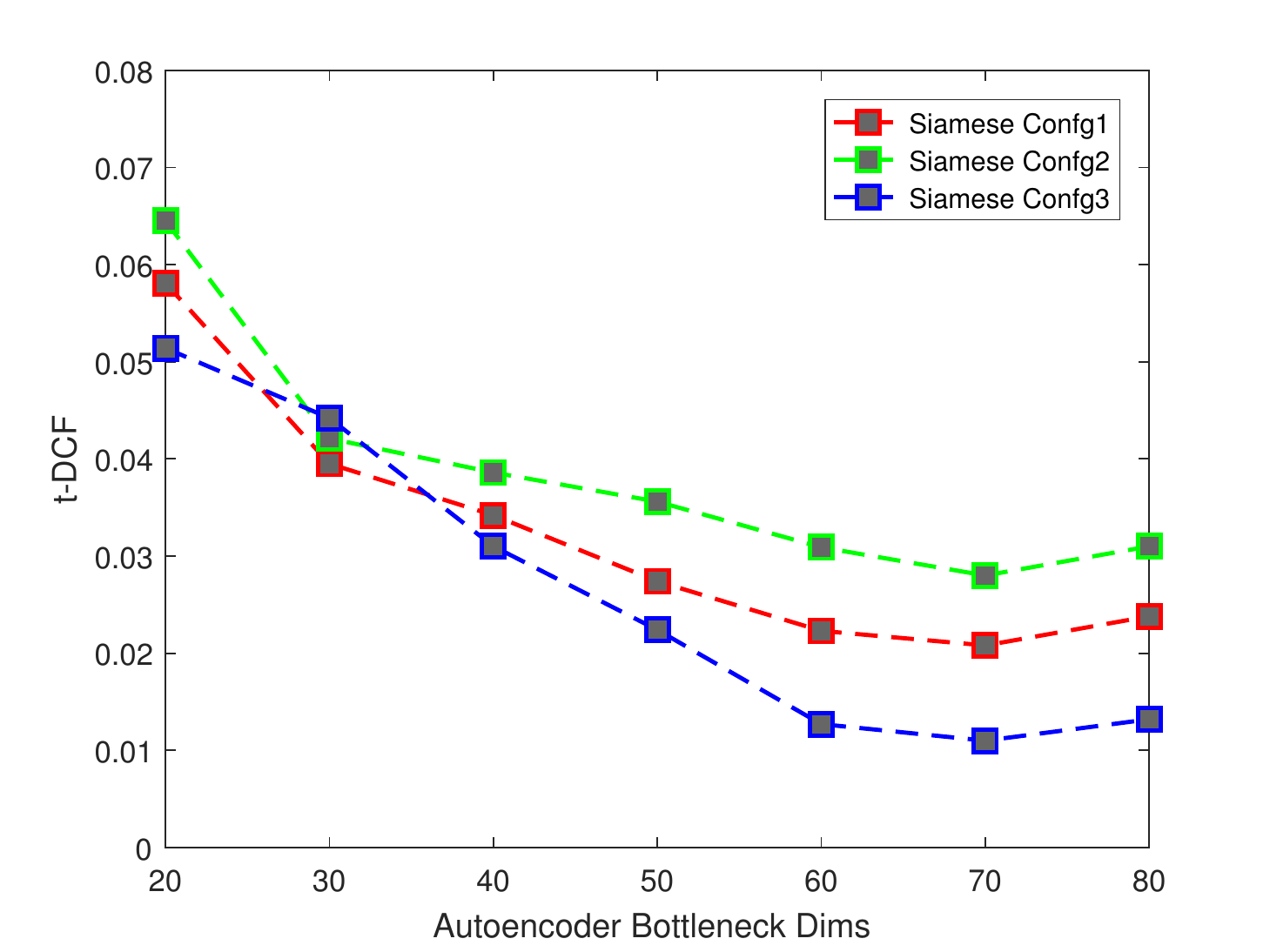}
%\vspace{-0.4cm}
\caption{Effects of dimension reduction by autoencoder on t-DCF for evaluation set with CQCC vector size of 90 on the proposed system.}
%\vspace{-0.2cm}
\label{fig:CQCC-AE-t-DCF}
\end{figure}

Fig. \ref{fig:CQCC-AE-EER} and Fig. \ref{fig:CQCC-AE-t-DCF} show the effects of using the autoencoder in our proposed system. Accordingly, we used the autoencoder to reduce the dimensions of the 90-dimensional CQCC. As shown in Figs. \ref{fig:CQCC-AE-EER} and \ref{fig:CQCC-AE-t-DCF}, best results are obtained from the 70-dimensional bottleneck of autoencoder for each Siamese configuration. Similar to previous results, the best results belong to Siamese Config. 3 that outperforms the baseline system by 10.72\% and 0.2344 in terms of EER and t-DCF, respectively.
\\
\textbf{Discussion:} Figs. \ref{fig:CQCC-AE-EER} and \ref{fig:CQCC-AE-t-DCF} show the same results for three different configurations, which were analyzed in the previous discussion. But it shows that the better performance for the proposed approach is given for higher bottleneck features. The purpose of using autoencoder is to consider the impact of noise, but attaching too much importance to the noise effect is associated with ignoring valuable information which leads to increment in EER. Therefore, there is a trade-off, as much as noise is considered, information is also missed and vice-versa. Consequently, there is an optimum value where most of the noises are considered and informative values are still preserved, which is 70 in this case.


Table \ref{tbl:AE-results} summarizes the best results obtained by different configurations of the proposed systems and baseline systems.

\begin{table}[ht]
\begin{center}
\caption {Comparison of the best results obtained by different Siamese networks configurations with baseline systems.(CQCC vector size = 90, AutoEncoder bottleneck size = 70)}

\label{tbl:AE-results}
\begin{tabular}{|p{0.8cm}|p{0.7cm}|c|p{0.2cm}|p{0.7cm}|p{0.2cm}|p{0.7cm}|}\cline{4-7}
\multicolumn{3}{c}{} & \multicolumn{2}{|c|}{Dev.} & \multicolumn{2}{c|}{Eval.}\\ \hline
\multicolumn{3}{|c|}{System} & \multicolumn{1}{c|}{EER\%}& \begin{scriptsize}t-DCF\end{scriptsize} & \multicolumn{1}{c|}{EER\%}&\begin{scriptsize}t-DCF\end{scriptsize}\\ \hline\hline
\multicolumn{3}{|c|}{Baseline 1 (LFCC + GMM)} & 11.96 & 0.2554  & 13.54 & 0.3017 \\ \hline
\multicolumn{3}{|c|}{Baseline 2 (CQCC + GMM)} & \textbf{9.87} & \textbf{0.1953}  &\textbf{11.04} & \textbf{0.2454} \\ \hline\hline

\multirow{6}{*}{} & \multirow{2}{*}{Conf. 1} & CQCC & 1.98 &  0.1529 & 3.27& 0.0745 \\ \cline{3-7}
{Siamese} & {} & CQCC + AE & \textbf{0.23} & \textbf{0.0027} & \textbf{0.80} & \textbf{0.0208} \\ \cline{2-7}
{Network} & \multirow{2}{*}{Conf. 2} & CQCC & 2.40 & 0.0448 & 4.13& 0.1086 \\ \cline{3-7}
{Configs.} & {} & CQCC + AE & \textbf{0.29}  & \textbf{0.0196} & \textbf{0.91} & \textbf{0.0278} \\ \cline{2-7}
{} & \multirow{2}{*}{Conf. 3} & CQCC & 1.77 & 0.0212 & 3.02 & 0.0627 \\ \cline{3-7}
{} & {} & CQCC + AE &\textbf{0.00} & \textbf{0.0041} & \textbf{0.62} & \textbf{0.0110} \\ \hline
%\multicolumn{2}{|c|}{System} & %\multicolumn{1}{|c}{EER\%}&t-DCF \\ \hline\hline
%\multicolumn{2}{|c|}{Baseline 1 (CQCC + GMM)} & \textbf{11.04} & \textbf{0.2454} \\ \hline
%\multicolumn{2}{|c|}{Baseline 2 (LFCC + GMM)}& 13.54 & 0.3017 \\ \hline\hline
%\multirow{2}{*}{Siamese Network Config1} & CQCC Vector Size 90 & 3.27& 0.0745 \\ \cline{2-4}
%{} & CQCC Vector Size 90 + 70-bottleneck AE & \textbf{0.80} & \textbf{0.0208} \\ \hline\hline
%\multirow{2}{*}{Siamese Network Config2} & CQCC Vector Size 90 & 4.13& 0.1086 \\ \cline{2-4}
%{} & CQCC Vector Size 90 + 70-bottleneck AE & \textbf{0.91} & \textbf{0.0278} \\ \hline\hline
%\multirow{2}{*}{Siamese Network Config3} & CQCC Vector Size 90 & 3.02& 0.0627 \\ \cline{2-4}
%{} & CQCC Vector Size 90 + 70-bottleneck AE & \textbf{0.62} & %\textbf{0.0110} \\ \hline
\end{tabular}
\vspace{-6mm}
\end{center}
\end{table}

\subsection{Performance With Different Training Data sizes}
\begin{figure}[H]
\centering
\includegraphics[width=0.45\textwidth]{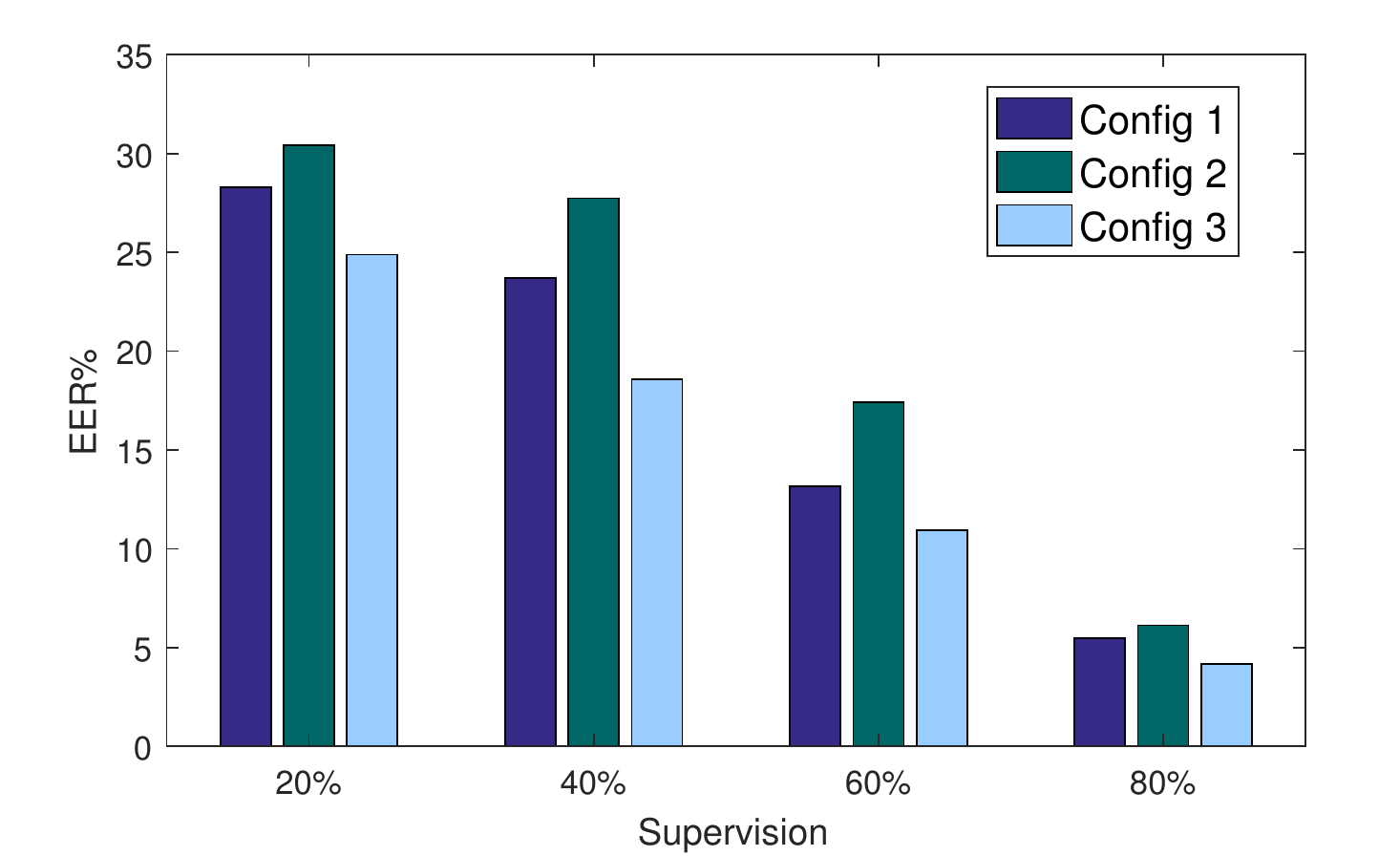}
%\vspace{-0.4cm}
\caption{The Effects of Using Different Sizes of the training set on the EER of the best configuration of Proposed System (CNN Config. 3, 90-dimensional CQCC, 70-dimensional AE bottleneck).}
%\vspace{-0.2cm}
\label{fig:size-eer}
\end{figure}

\begin{figure}[H]
\centering
\includegraphics[width=0.45\textwidth]{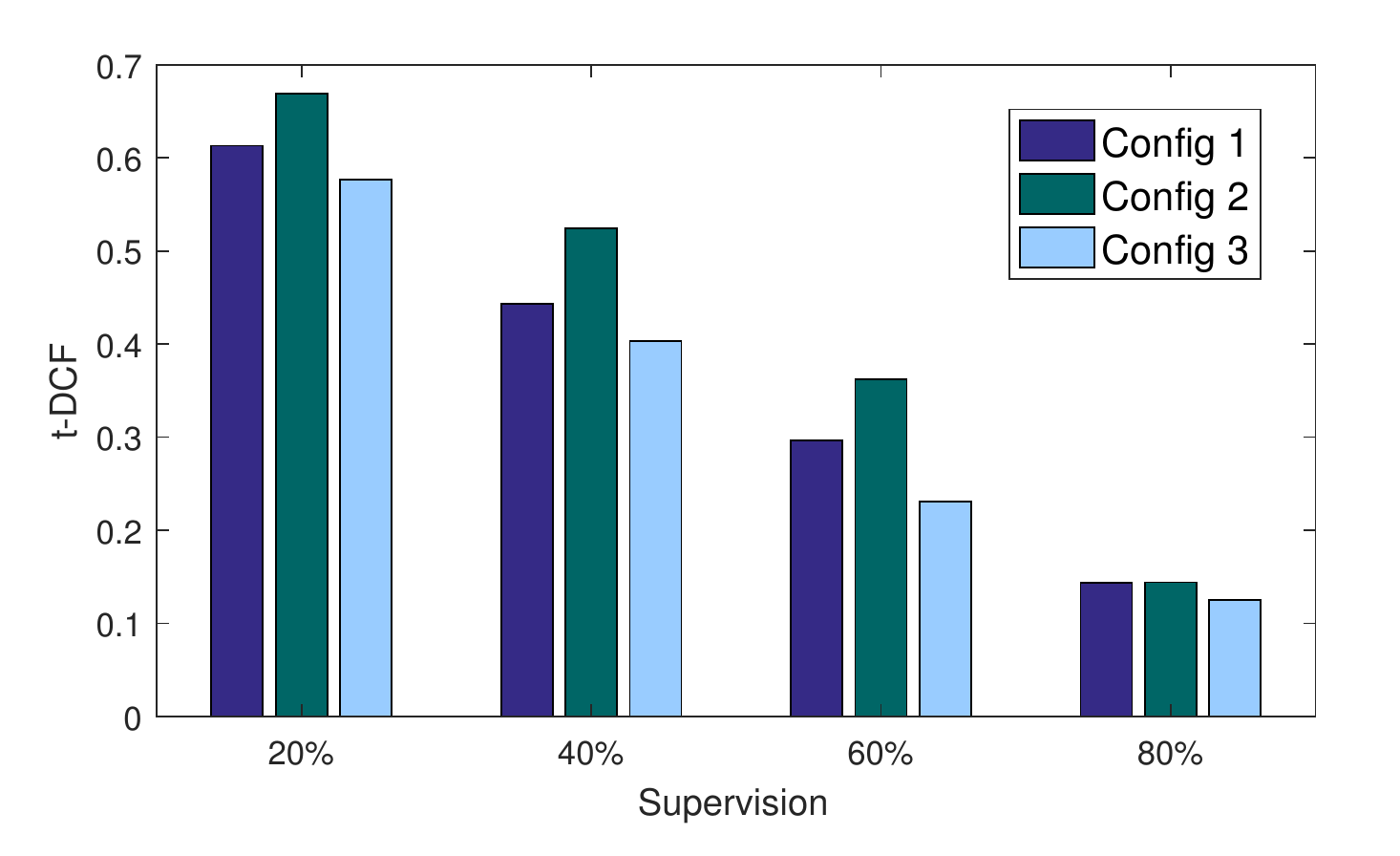}
%\vspace{-0.4cm}
\caption{The Effects of Using Different Sizes of the training set on the t-DCF of the best configuration of Proposed System (CNN Config. 3, 90-dimensional CQCC, 70-dimensional AE bottleneck).}
%\vspace{-0.2cm}
\label{fig:size-t-DCF}
\end{figure}

In order to measure the performance of our proposed system in face of different training data, we used different sizes of the training set to train the system. In this regard, we randomly divided the training set into five equal folds without considering the gender or number of speakers. In each step, the training set was raised fold by fold. For each fold, the system is trained similar to what we do in the training phase. Fig. \ref{fig:size-eer} and Fig. \ref{fig:size-t-DCF} show the effects of using different sizes of the training set. As expected, the results were improved by increasing the volume of the training data. It is noteworthy that our proposed system could achieve better results than the baseline system on the evaluation set using only 60\% of the training data. By 80\% of data as the training set, the proposed approach outperforms the baseline system.
\\
\textbf{Discussion:} The results for this part demonstrate better outcomes for more amount of data which is kind of obvious. Increasing training data helps the classifier to figure out the data model which leads to better prediction of test samples, however, considering 60\% of whole data as a training set provides comparable results to the baseline system, and 80\% of the whole data gives a significant improvement over the baseline system. Another important subject is that the proposed system demonstrates impressive results using less training data. This makes the proposed approach more robust to missing data. It means the proposed system can be applied to tasks with less training data and since data scarcity is a serious issue, it is a strong advantage for the proposed approach. 

\subsection{Effect of Siamese Networks on Performance}

To evaluate the performance of the Siamese networks, the results for employed Siamese networks and single CNN are compared at table \ref{tbl:siamese-vs-cnn}. The hyper-parameters of CQCC and autoencoder are tuned based on the best performance of the system. Therefore, the dimension of CQCC is fixed on 90, and for the autoencoder we used bottleneck with a size of 70. Three different configurations are used for each case. The results indicate that Siamese networks are effective at improving the results compared with single CNN.
\begin{table}[htbp]
\begin{center}
\caption {Effect of employed Siamse network on the performance (CQCC vector size = 90,  BOTTLENECK SIZE= 70).}
\label{tbl:siamese-vs-cnn}
\begin{tabular}{|c|c|c|c|c|c|}\cline{3-6}
\multicolumn{1}{c}{}  & \multicolumn{1}{c}{} & \multicolumn{2}{|c|}{Siamese Network} & \multicolumn{2}{|c|}{CNN} \\ \hline
\multicolumn{1}{|c|}{Set} & \multicolumn{1}{c}{Configurations} & \multicolumn{1}{|c|}{EER} & \multicolumn{1}{|c|}{t-DCF} & \multicolumn{1}{|c|}{EER} & \multicolumn{1}{|c|}{t-DCF} \\ \hline

\multirow{3}{*}{Dev.} & Conf. 1 & 0.23 & 0.0027 & 0.94 & 0.0212 \\ \cline{2-6}
{}& Conf. 2 & 0.29 & 0.0196 & 1.21 & 0.0403 \\ \cline{2-6}
{}& Conf. 3 & \textbf{0.00} & \textbf{0.0041} &  \textbf{0.66} &  \textbf{0.0182} \\ \hline \hline

\multirow{3}{*}{Eval.} & Conf. 1 &  0.80 & 0.0208 & 3.25 & 0.1339 \\ \cline{2-6}
{} & Conf. 2 & 0.91 & 0.0278 & 5.11 & 0.2726 \\ \cline{2-6}
{} & Conf. 3 & \textbf{0.62} & \textbf{0.0110} & \textbf{3.05} & \textbf{0.1406} \\ \hline

%Siamese Config1 &\multicolumn{1}{c}{} & \multicolumn{1}{c|}{} & Siamese Config2 & \multicolumn{1}{c}{} & \multicolumn{1}{c|}{} & Siamese Config3& \multicolumn{1}{c}{} & \multicolumn{1}{c|}{} \\ \cline{1-1}\cline{4-4}\cline{7-7}
%CQCC Dimensions &\multicolumn{1}{c}{EER\%} & \multicolumn{1}{c|}{t-DCF} & CQCC Dimensions &\multicolumn{1}{c}{EER\%} & \multicolumn{1}{c|}{t-DCF} & CQCC Dimensions &\multicolumn{1}{c}{EER\%} & \multicolumn{1}{c|}{t-DCF} \\ \hline\hline
%30 & 6.80 & 0.1550 & 30 & 7.12 & 0.1649 & 30 & 5.66 & .1371 \\ \hline
%60 & 4.19 & 0.1072 & 60 & 5.57 & 0.1323 & 60 & 3.78 & 0.0883 \\ \hline
%90 & \textbf{3.27} & \textbf{0.0745} & 90 & \textbf{4.13} & \textbf{0.1086} & 90 & \textbf{3.02} & \textbf{0.0627} \\ \hline
\end{tabular}
\vspace{-6mm}
\end{center}
\end{table}
\\

\textbf{Discussion:} Siamese networks have shown promising results in different classification tasks and their application was restricted to classification tasks with an unlimited number of classes. Hence, their effectiveness was overlooked in such tasks. However, as Table \ref{tbl:siamese-vs-cnn} indicates, the Siamese networks improve the performance of the system. Although CNN yields satisfactory results, it does show weaknesses in  stability. In other words, the system performance varies with different configurations, and the results for the evaluation set are much worse than the results for the development set. This is mostly because both classes are trained on one network which somehow reveals some evidence of over-fitting. In addition, as a result of the network discriminates between different samples, the network has difficulty in handling the different inputs, especially in case of using average-pooling which removes the noises. On the other hand, for Siamese networks, the difference of generated output is learnt and this is one step in addition to normal classification. In this case, even with the average for pooling, the model is learnt through the final layer. However, the average pooling would remove the noises and leads to a drop in performance of classification.

\section{Conclusion}
\label{sec:conclusion}
We proposed a novel replay spoofing countermeasure system for ASVs based on physical access ASVspoof 2019 dataset. In this study, different configurations of CNNs in the structure of Siamese Network were investigated for the purpose of classification. Moreover, the autoencoder is employed to resolve the dispersion problem of well-known CQCC features. The experimental results confirmed the high efficiency of Siamese Network for spoofing detection. Additionally, autoencoders could significantly consider and utilize noise and eliminate redundant and irrelevant information of CQCC features, resulting in outperforming the baseline system. Besides, the results show that the proposed method could achieve comparable performance using a smaller amount of training set (only 60\% training set) which reflects the effectiveness and the scalability of our system.

For future work, an integrated framework can be proposed to identify each of four general attacks on ASVspoof systems and apply appropriate countermeasures regarding the attack. It is a necessary task since every ASV system is always facing other types of attacks and has to be able to confront each type of attack.

\bibliographystyle{IEEEtran}
\bibliography{paper.bib}

\begin{thebibliography}{10}
\providecommand{\url}[1]{#1}
\csname url@samestyle\endcsname
\providecommand{\newblock}{\relax}
\providecommand{\bibinfo}[2]{#2}
\providecommand{\BIBentrySTDinterwordspacing}{\spaceskip=0pt\relax}
\providecommand{\BIBentryALTinterwordstretchfactor}{4}
\providecommand{\BIBentryALTinterwordspacing}{\spaceskip=\fontdimen2\font plus
\BIBentryALTinterwordstretchfactor\fontdimen3\font minus
  \fontdimen4\font\relax}
\providecommand{\BIBforeignlanguage}[2]{{%
\expandafter\ifx\csname l@#1\endcsname\relax
\typeout{** WARNING: IEEEtran.bst: No hyphenation pattern has been}%
\typeout{** loaded for the language `#1'. Using the pattern for}%
\typeout{** the default language instead.}%
\else
\language=\csname l@#1\endcsname
\fi
#2}}
\providecommand{\BIBdecl}{\relax}
\BIBdecl

\bibitem{Jain2006}
A.~K. Jain, A.~Ross, and S.~Pankanti, ``{Biometrics: a tool for information
  security},'' \emph{IEEE transactions on information forensics and security},
  vol.~1, no.~2, pp. 125--143, 2006.

\bibitem{Kinnunen2010}
T.~Kinnunen and H.~Li, ``{An overview of text-independent speaker recognition:
  From features to supervectors},'' \emph{Speech communication}, vol.~52,
  no.~1, pp. 12--40, 2010.

\bibitem{Amin2014}
T.~B. Amin, P.~Marziliano, and J.~S. German, ``{Glottal and vocal tract
  characteristics of voice impersonators},'' \emph{IEEE Transactions on
  Multimedia}, vol.~16, no.~3, pp. 668--678, 2014.

\bibitem{Masuko2000}
T.~Masuko, K.~Tokuda, and T.~Kobayashi, ``{Imposture using synthetic speech
  against speaker verification based on spectrum and pitch},'' in \emph{Sixth
  International Conference on Spoken Language Processing}, 2000.

\bibitem{Lee2013}
K.~A. Lee, B.~Ma, and H.~Li, ``{Speaker verification makes its debut in
  smartphone},'' \emph{IEEE signal processing society speech and language
  technical committee newsletter}, 2013.

\bibitem{Lau2004}
Y.~W. Lau, M.~Wagner, and D.~Tran, ``{Vulnerability of speaker verification to
  voice mimicking},'' in \emph{Proceedings of 2004 International Symposium on
  Intelligent Multimedia, Video and Speech Processing, 2004.}\hskip 1em plus
  0.5em minus 0.4em\relax IEEE, 2004, pp. 145--148.

\bibitem{Eriksson1997}
A.~Eriksson and P.~Wretling, ``{How flexible is the human voice?-A case study
  of mimicry},'' in \emph{Fifth European Conference on Speech Communication and
  Technology}, 1997.

\bibitem{Stylianou2009}
Y.~Stylianou, ``{Voice transformation: a survey},'' in \emph{2009 IEEE
  International Conference on Acoustics, Speech and Signal Processing}.\hskip
  1em plus 0.5em minus 0.4em\relax IEEE, 2009, pp. 3585--3588.

\bibitem{Evans2015}
N.~Evans, F.~Alegre, Z.~Wu, and T.~Kinnunen, ``{Anti-spoofing, voice
  conversion},'' \emph{Encyclopedia of Biometrics}, pp. 115--122, 2015.

\bibitem{Lindberg1999}
J.~Lindberg and M.~Blomberg, ``{Vulnerability in speaker verification-a study
  of technical impostor techniques},'' in \emph{Sixth European Conference on
  Speech Communication and Technology}, 1999.

\bibitem{Villalba2010}
J.~Villalba and E.~Lleida, ``{Speaker verification performance degradation
  against spoofing and tampering attacks},'' in \emph{FALA workshop}, 2010, pp.
  131--134.

\bibitem{Kaavya2016}
S.~Kaavya, V.~Sethu, P.~N. Le, and E.~Ambikairajah, ``{Investigation of
  sub-band discriminative information between spoofed and genuine speech},'' in
  \emph{Interspeech}, 2016, pp. 1710--1714.

\bibitem{Kaavya2018}
S.~Kaavya, V.~Sethu, and E.~Ambikairajah, ``{Deep Siamese Architecture Based
  Replay Detection for Secure Voice Biometric},'' in \emph{Interspeech}, 2018,
  pp. 671--675.

\bibitem{bell2015learning}
S.~Bell and K.~Bala, ``Learning visual similarity for product design with
  convolutional neural networks,'' \emph{ACM Transactions on Graphics (TOG)},
  vol.~34, no.~4, p.~98, 2015.

\bibitem{Adiban2017}
M.~Adiban, H.~Sameti, N.~Maghsoodi, and S.~Shahsavari, ``{SUT System
  Description for Anti-Spoofing 2017 Challenge},'' in \emph{Proceedings of the
  29th Conference on Computational Linguistics and Speech Processing (ROCLING
  2017)}, 2017, pp. 264--275.

\bibitem{Yin2015}
S.~Yin, C.~Liu, Z.~Zhang, Y.~Lin, D.~Wang, J.~Tejedor, T.~F. Zheng, and Y.~Li,
  ``Noisy training for deep neural networks in speech recognition,''
  \emph{EURASIP Journal on Audio, Speech, and Music Processing}, vol. 2015,
  no.~1, pp. 1--14, 2015.

\bibitem{Sun2016}
M.~Sun, X.~Zhang, and T.~F. Zheng, ``{Unseen noise estimation using separable
  deep auto encoder for speech enhancement},'' \emph{IEEE/ACM Transactions on
  Audio, Speech and Language Processing (TASLP)}, vol.~24, no.~1, pp. 93--104,
  2016.

\bibitem{hsu2017unsupervised}
W.-N. Hsu, Y.~Zhang, and J.~Glass, ``Unsupervised domain adaptation for robust
  speech recognition via variational autoencoder-based data augmentation,'' in
  \emph{2017 IEEE Automatic Speech Recognition and Understanding Workshop
  (ASRU)}.\hskip 1em plus 0.5em minus 0.4em\relax IEEE, 2017, pp. 16--23.

\bibitem{Villalba2011}
J.~Villalba and E.~Lleida, ``{Detecting replay attacks from far-field
  recordings on speaker verification systems},'' in \emph{European Workshop on
  Biometrics and Identity Management}.\hskip 1em plus 0.5em minus 0.4em\relax
  Springer, 2011, pp. 274--285.

\bibitem{Ji2017}
Z.~Ji, Z.-Y. Li, P.~Li, M.~An, S.~Gao, D.~Wu, and F.~Zhao, ``{Ensemble Learning
  for Countermeasure of Audio Replay Spoofing Attack in ASVspoof2017.}'' in
  \emph{Interspeech}, 2017, pp. 87--91.

\bibitem{Wu2014}
Z.~Wu, S.~Gao, E.~S. Cling, and H.~Li, ``{A study on replay attack and
  anti-spoofing for text-dependent speaker verification},'' in \emph{In Signal
  and Information Processing Association Annual Summit and Conference
  (APSIPA)}, 2014, pp. 1--5.

\bibitem{Shang2010}
W.~Shang and M.~Stevenson, ``{Score normalization in playback attack
  detection},'' in \emph{2010 IEEE International Conference on Acoustics,
  Speech and Signal Processing}.\hskip 1em plus 0.5em minus 0.4em\relax IEEE,
  2010, pp. 1678--1681.

\bibitem{kinnunen2012vulnerability}
T.~Kinnunen, Z.-Z. Wu, K.~A. Lee, F.~Sedlak, E.~S. Chng, and H.~Li,
  ``Vulnerability of speaker verification systems against voice conversion
  spoofing attacks: The case of telephone speech,'' in \emph{2012 IEEE
  International Conference on Acoustics, Speech and Signal Processing
  (ICASSP)}.\hskip 1em plus 0.5em minus 0.4em\relax IEEE, 2012, pp. 4401--4404.

\bibitem{Wang2011}
Z.-F. Wang, G.~Wei, and Q.-H. He, ``{Channel pattern noise based playback
  attack detection algorithm for speaker recognition},'' in \emph{2011
  International conference on machine learning and cybernetics}, vol.~4.\hskip
  1em plus 0.5em minus 0.4em\relax IEEE, 2011, pp. 1708--1713.

\bibitem{Shim2018}
H.-J. Shim, J.-W. Jung, H.-S. Heo, S.-H. Yoon, and H.-J. Yu, ``{Replay spoofing
  detection system for automatic speaker verification using multi-task learning
  of noise classes},'' in \emph{2018 Conference on Technologies and
  Applications of Artificial Intelligence (TAAI)}.\hskip 1em plus 0.5em minus
  0.4em\relax IEEE, 2018, pp. 172--176.

\bibitem{Alam2018}
M.~J. Alam, G.~Bhattacharya, and P.~Kenny, ``{Boosting the performance of
  spoofing detection systems on replay attacks using q-logarithm domain feature
  normalization},'' in \emph{Proc. Odyssey 2018 The Speaker and Language
  Recognition Workshop}, 2018, pp. 393--398.

\bibitem{Lavrentyeva2017}
G.~Lavrentyeva, S.~Novoselov, E.~Malykh, A.~Kozlov, O.~Kudashev, and
  V.~Shchemelinin, ``Audio replay attack detection with deep learning
  frameworks.'' in \emph{Interspeech}, 2017, pp. 82--86.

\bibitem{Wickramasinghe2018}
B.~Wickramasinghe, S.~Irtza, E.~Ambikairajah, and J.~Epps, ``{Frequency Domain
  Linear Prediction Features for Replay Spoofing Attack Detection},'' in
  \emph{Interspeech}, 2018, pp. 661--665.

\bibitem{Sailor2018}
H.~Sailor, M.~Kamble, and H.~Patil, ``{Auditory filterbank learning for
  temporal modulation features in replay spoof speech detection},'' in
  \emph{Interspeech}, 2018, pp. 666--670.

\bibitem{cai2019dku}
W.~Cai, H.~Wu, D.~Cai, and M.~Li, ``The dku replay detection system for the
  asvspoof 2019 challenge: On data augmentation, feature representation,
  classification, and fusion,'' \emph{arXiv preprint arXiv:1907.02663}, 2019.

\bibitem{li2019anti}
R.~Li, M.~Zhao, Z.~Li, L.~Li, and Q.~Hong, ``Anti-spoofing speaker verification
  system with multi-feature integration and multi-task learning,'' \emph{Proc.
  Interspeech 2019}, pp. 1048--1052, 2019.

\bibitem{alzantot2019deep}
M.~Alzantot, Z.~Wang, and M.~B. Srivastava, ``Deep residual neural networks for
  audio spoofing detection,'' \emph{arXiv preprint arXiv:1907.00501}, 2019.

\bibitem{das2019long}
R.~K. Das, J.~Yang, and H.~Li, ``Long range acoustic features for spoofed
  speech detection,'' in \emph{20th Annual Conference of the International
  Speech Communication Association (INTERSPEECH)}, 2019.

\bibitem{lavrentyeva2019stc}
G.~Lavrentyeva, S.~Novoselov, A.~Tseren, M.~Volkova, A.~Gorlanov, and
  A.~Kozlov, ``Stc antispoofing systems for the asvspoof2019 challenge,''
  \emph{arXiv preprint arXiv:1904.05576}, 2019.

\bibitem{chettri2019ensemble}
B.~Chettri, D.~Stoller, V.~Morfi, M.~A.~M. Ram{\'\i}rez, E.~Benetos, and B.~L.
  Sturm, ``Ensemble models for spoofing detection in automatic speaker
  verification,'' \emph{arXiv preprint arXiv:1904.04589}, 2019.

\bibitem{Todisco2016}
M.~Todisco, H.~Delgado, and N.~W.~D. Evans, ``{Articulation Rate Filtering of
  CQCC Features for Automatic Speaker Verification.}'' in \emph{Interspeech},
  2016, pp. 3628--3632.

\bibitem{Brown1991}
J.~C. Brown, ``{Calculation of a constant Q spectral transform},'' \emph{The
  Journal of the Acoustical Society of America}, vol.~89, no.~1, pp. 425--434,
  1991.

\bibitem{Schorkhuber2010}
C.~Sch{\"{o}}rkhuber and A.~Klapuri, ``{Constant-Q transform toolbox for music
  processing},'' in \emph{7th Sound and Music Computing Conference, Barcelona,
  Spain}, 2010, pp. 3--64.

\bibitem{Todisco2016a}
M.~Todisco, H.~Delgado, and N.~Evans, ``{A new feature for automatic speaker
  verification anti-spoofing: Constant Q cepstral coefficients},'' in
  \emph{Speaker Odyssey Workshop, Bilbao, Spain}, vol.~25, 2016, pp. 249--252.

\bibitem{ChopraS.HadsellR.2005}
L.~Y. {Chopra, S., Hadsell, R.}, ``{Learning a similarity metric
  discriminatively, with application to face verification},'' in \emph{Proc. of
  Computer Vision and Pattern Recognition. IEEE,}.\hskip 1em plus 0.5em minus
  0.4em\relax Proc. of Computer Vision and Pattern Recognition. IEEE, 2005, pp.
  539--546.

\bibitem{BromleyJ.BentzJ.W.ButtouL.GuyonI.LeCunY.MooreC.SackingerE.Shah1993}
R.~{Bromley, J., Bentz, J. W., Buttou, L., Guyon, I., LeCun, Y., Moore, C.,
  Sackinger, E., Shah}, ``{Signature verification using a siamese time delay
  neural network},'' \emph{International Journal of Pattern Recognition and
  Artificial Intelligence}, vol.~7, pp. 669--688, 1993.

\bibitem{Yang2014LearnCN}
J.~Yang, Z.~Lei, and S.~Z. Li, ``Learn convolutional neural network for face
  anti-spoofing,'' \emph{arXiv preprint arXiv:1408.5601}, 2014.

\bibitem{ma2018deep}
Z.~Ma, Y.~Ding, B.~Li, and X.~Yuan, ``Deep cnns with robust lbp guiding pooling
  for face recognition,'' \emph{Sensors}, vol.~18, no.~11, p. 3876, 2018.

\bibitem{wu2015lightened}
X.~Wu, R.~He, and Z.~Sun, ``A lightened cnn for deep face representation,''
  \emph{arXiv preprint arXiv:1511.02683}, vol.~4, no.~8, 2015.

\bibitem{Greff2015}
K.~Greff, ``{Highway Networks},'' \emph{arXiv preprint arXiv:1505.00387}, 2015.

\bibitem{Computing2017}
K.~Ahrabian and B.~Babaali, ``Usage of autoencoders and siamese networks for
  online handwritten signature verification,'' \emph{Neural Computing and
  Applications}, pp. 1--14, 2018.

\bibitem{Todisco2019}
M.~Todisco, X.~Wang, V.~Vestman, M.~Sahidullah, H.~Delgado, A.~Nautsch,
  J.~Yamagishi, N.~Evans, T.~Kinnunen, and K.~A. Lee, ``{ASVspoof 2019: Future
  Horizons in Spoofed and Fake Audio Detection},'' \emph{arXiv preprint
  arXiv:1904.05441}, 2019.

\bibitem{Wu2015}
Z.~Wu, T.~Kinnunen, N.~Evans, J.~Yamagishi, C.~Hanil{\c{c}}i, M.~Sahidullah,
  and A.~Sizov, ``{ASVspoof 2015: the first automatic speaker verification
  spoofing and countermeasures challenge},'' in \emph{Sixteenth Annual
  Conference of the International Speech Communication Association}, 2015.

\bibitem{Kinnunen2017}
T.~Kinnunen, M.~Todisco, N.~Evans, J.~Yamagishi, and K.~A. Lee, ``{The ASVspoof
  2017 Challenge : Assessing the Limits of Replay Spoofing Attack Detection},''
  in \emph{Interspeech}, 2017, pp. 2--6.

\bibitem{Kinnunen2018}
T.~Kinnunen, K.~A. Lee, H.~Delgado, N.~Evans, M.~Todisco, M.~Sahidullah,
  J.~Yamagishi, and D.~A. Reynolds, ``{t-DCF: a detection cost function for the
  tandem assessment of spoofing countermeasures and automatic speaker
  verification},'' \emph{arXiv preprint arXiv:1804.09618}, 2018.

\bibitem{Sahidullah2015}
M.~Sahidullah, T.~Kinnunen, and C.~Hanil, ``{A comparison of features for
  synthetic speech detection},'' in \emph{Interspeech}, 2015, pp. 2087--2091.

\bibitem{Kinnunen2016}
T.~Kinnunen, N.~Evans, J.~Yamagishi, K.~A. Lee, M.~Sahidullah, M.~Todisco, and
  H.~Delgado, ``{ASVspoof 2019: Automatic Speaker Verification Spoofing and
  Countermeasures Challenge Evaluation Plan},''
  \emph{http://www.asvspoof.org/asvspoof2019{\_}evaluation{\_}plan}, 2019.

\end{thebibliography}

\end{document}